\documentclass[prl,showpacs,byrevtex]{revtex4}
\input epsf.tex
\input epsf
\usepackage{psfig}
\usepackage{epsf}
\usepackage{bm}
\def\ba{\begin{eqnarray}}
\def\ea{\end{eqnarray}}
\def\br{\begin{array}}
\def\er{\end{array}}
\def\be{\begin{equation}}
\def\ee{\end{equation}}
\def\ol{\overline}
\def\gf{SO(10)\times S_4} 

\def\op{\oplus}
\def\dr{\Delta_R^o}
\def\drb{\overline {\Delta_R^o}}
\def\sg{\sigma}
\def\sgb{\overline{\sigma}}
\def\Sg0{\Sigma_0}
\def\Sg1{\Sigma_1}
\def\Sg2{\Sigma_2}
\def\l{\lambda}
\def\Dl{\Delta_L}

\def\Dr{\Delta_R}
   
\def\al{\alpha}
\def\al0{\alpha_0}
\def\al1{\alpha_1}
\def\al2{\alpha_2}
\def\al3{\alpha_3}
\def\al4{\alpha_4}
\def\al5{\alpha_5}
\def\bt0{\beta_0}
\def\bt1{\beta_1}
\def\bt2{\beta_2}
\def\bt3{\beta_3}
\def\bt4{\beta_4}
\def\bt5{\beta_5}
\def\gm0{\gamma_0}
\def\gm1{\gamma_1}
\def\gm2{\gamma_2}
\def\dl0{\delta_0}
\def\dl1{\delta_1}
\def\dl2{\delta_2}

\begin{document}
\title{Intermediate  left-right gauge symmetry, unification of
  couplings and 
fermion masses in supersymmetric $\mathbf{SO(10)\times S_4}$ }
\author{M. K. Parida}
\email{mkparida@iopb.res.in}
\affiliation{National Institute for Science  Education and Research\\
Institute of Physics Campus, Sachivalaya Marg, Bhubaneswar 751005, India}
\begin{abstract}
If left-right gauge theory,  $SU(2)_L\times SU(2)_R\times
U(1)_{B-L}\times SU(3)_C(g_L=g_R)(\equiv G_{2213})$, occurs as an
intermediate symmetry in a grand unified theory then, apart from other
advantages, it is possible to obtain the seesaw scale necessary to
understand small neutrino masses with 
 Majorana coupling of order unity.
 Barring threshold or
non-renormalizable gravitational  effects at the GUT scale, or the assumed  presence of additional light scalar
particles of unprescribed origin, all other  attempts to achieve  manifest one-loop gauge coupling unification in SUSY  $SO(10)$ 
with such  intermediate 
symmetry have not been successful
so far. 
Attributing this failure to lack of flavor symmetry in the grand 
unified theory, we show how the spontaneous symmetry breaking of 
$SO(10)\times S_4$  leads  to such left-right intermediate breaking scale
extending over the range $M_R \simeq 5\times 10^{9}$ GeV to $10^{15}$ GeV.
All the charged fermion masses  are fitted at the intermediate see-saw scale, 
$M_N\simeq M_R \simeq 4\times 10^{13}$ GeV which is obtained with
 Majorana coupling $f_0\simeq 1$. Using type I seesaw and a constrained parametrisation 
in which CP-violation originates only from the quark sector, in addition to other predictions 
made in the  neutrino sector, the reactor mixing angle is found to be 
$\theta_{13} \simeq 3^{\circ} - 5^{\circ}$ which is in the range accessible to ongoing experiments. The leptonic Dirac phase turns out to be $\delta \simeq 
2.9- 3.1$ radians with the predicted values of Jarlskog invariant $J_{CP}\simeq
2.95 \times 10^{-5} -  10^{-3}$.

\end{abstract}
\date{ 26 April 2008}
\maketitle
\par

\section{I. Introduction}\label{section1}
   $SO(10)$ grand unified theory \cite{so10} has a number of
 attractive features which have resulted in recent surge of investigations including applications to fermion masses and mixings \cite{rnm5}.  It unifies all fermions of one generation plus the right-handed
 neutrino into one spinorial representation ${\bf {16}}$. With D-parity as an
 element of gauge transformation \cite{cmp}, it naturally restores left-right
 and CP 
 symmetries at the GUT scale and thus, it can provide a spontaneous
origin of P and CP-violations \cite{rnm1}.
It embodies quark-lepton unification with high predictive power in the
 fermionic sector \cite{pati,mpr}.
Through its Higgs representations ${\bf {10}}$ 
and
 ${\bf {\overline {126}}}$ or ${\bf {\overline {16}}}$, 
 it has the potentialities for intermediate
$SU(2)_R \times U(1)_{B-L}$ breaking, generation of small Majorana neutrino 
masses through type-I, type-II, and type III see-saw mechanisms \cite{seesaw, rnm2, barr}, 
explanation of large 
neutrino mixings through type-II see-saw dominance, accounting for dominant
 contributions to  charged fermion masses through the ${\bf 10}$-representation 
and providing the desired corrections to them through the weak-doublets in 
${\bf {\ol {126}}}$  and ${\bf {120}
}$ \cite {babu1,goh1,goh2,bajc1,berto,dutta, grimus} . 
 With natural R-parity conservation,
in addition to ensuring proton stability \cite{rnmp}, it predicts 
the lightest supersymmetric particle as a
stable
 dark matter candidate. 

All fermion masses and mixings, including very small masses and large
 mixings in the neutrino sector, have been shown to fit reasonably well 
if the right-handed neutrino mass scale is in the range $10^{13}$ GeV to $10^{14}$
GeV \cite{goh1, berto} .
Further, thermal
 leptogenesis explaining origin of matter through baryogenesis can be 
  implemented if the right handed neutrino masses are in
 intermediate range \cite{yana,ji}. Therefore, it would be interesting to obtain 
the right-handed mass scale  near  the intermediate breaking of $SU(2)_L\times SU(2)_R
\times SU(4)_C (\equiv G_{224}) \subset SO(10)$ or any of its subgroups
 like $G_{2213}$.
However, the   mass spectra analysis   in the minimal SUSY 
$SO(10)$ model \cite{bajc0} with
the Higgs representations $210\oplus 126 \oplus {\ol {126}}\oplus 10$ rules
out any possibility of intermediate gauge symmetry by predicting additional light scalars which are found to
disrupt gauge coupling unification \cite{fuku1,fuku2,bajc2}. It has been also noted that, even if these
additional light scalars are made naturally superheavy using a non-minimal Higgs
representation, leaving only the minimal light Higgs spectrum necessary to 
implement spontaneous symmetry breaking in the presence of supersymmetry and
R-parity conservation, there exists no intermediate scale through 
manifest one-loop 
unification of gauge couplings.

Recently attempts  have been made to obtain desired see-saw scales
and improved fits to  the fermion masses  
using different  types of mechanisms or by invoking GUTs in higher dimensions  
 \cite{goh3, grimus}.  
In addition, extensions of gauge symmetries by non-abelian flavor
groups like $S_3, S_4$, and $A_4$ have resulted in 
interesting consequences \cite{mpr, pakvasa, rnm3, lee, hagedorn,a4}
 including  high scale unification  of quark and lepton
mixings \cite{mpr}.

Since SUSY GUTs with a left-right
intermediate gauge symmetry  has many attractive features over super-grand desert models, in this paper we discuss a  novel procedure of constructing such a 
model
through an extension of the left-right gauge symmtry to 
$G_{2213}\times S_4$ and the corresponding  extension of the GUT 
symmetry to $\gf$ to encompass supersymmetric flavor
unification, grand unification, and R-parity conservation \cite{mkp0}.
We  find that when such a flavor symmetry is included, manifestly successful one-loop
gauge coupling unification occurs with the three gauge couplings of the 
left-right gauge theory attaining convergence to the
unification coupling at the GUT scale.   
 The intermediate
scale is predicted over a wide  range with $M_R \approx  5 \times 10^9$ GeV to $10^{15}$ GeV. The desired Higgs scalars necessary for the gauge coupling unification 
are found to be consistent with the  mass spectra analysis in the $\gf$  model.
 Although the contribution of three
fermion generations cancel out from  one-loop unification, flavor symmetry requires enlarged Higgs
spectrum which modifies the beta function coefficients  and  
the evolution of the gauge couplings leading to successful unification  in the presence of $G_{2213}\times S_4$ intermediate
 symmetry. In the second part of the paper, we fit the renormalization group~(RG)
extrapolated data on fermion masses, mixings, and phases at the intermediate see-saw scale $M_N \simeq M_R \simeq 10^{13}$ GeV and obtain successful  predictions in  the neutrino sector.

This paper is organized in the following manner. In Sec.2 we briefly review
problems associated with realization of an intermediate scale in SUSY
$SO(10)$. In Sec. III we discuss how the desired intermediate scale is achieved 
via left-right gauge  and $S_4$ flavor symmetries. Symmetry breaking of $\gf$ is discussed
in Sec.IV. In Sec.V we show how the desired light particle spectrum  
 is obtained from the  $\gf$ theory.
Fits to the fermion masses and mixings and model predictions in the neutrino sector are carried out in Sec.VI.
A brief summary of investigations made  and conclusions obtained
are stated in Sec.VII.

\section{II. Difficulties in R-Parity Conserving Left-Right  Intermediate Gauge Symmetry}\label{section2}

In this section we discuss briefly  the problems associated with obtaining 
a $SU(2)_L\times SU(2)_R\times U(1)_{B-L}\times  SU(3)_C(g_L=g_R)$($\equiv
G_{2213}$) intermediate gauge symmetry in R-parity conserving supersymmetric SO(10) grand
unified theory through  one-loop  unification of gauge couplings.

For this purpose,
we consider two-step breaking of $SO(10)$ to the minimal supersymmetric standard model(MSSM) through $G_{2213}$
intermediate gauge symmetry in the so called minimal grand unified theory,  

\begin{eqnarray} &&SO(10)
  ~\mathop{\longrightarrow}^{\underline{210}}_{M_U}~G_{2213}
\nonumber\\ &&\mathop{\longrightarrow}^{\underline {126}\oplus 
\underline {\overline{126}}}_{M_R}
~G_{213}~
\mathop{\longrightarrow}^{\underline{10}}_{M_Z}~U(1)_{em}\times SU(3)_C.
\label{chn}\end{eqnarray}
\par\noindent
The $G_{224}$ sub-multiplet $(1,1,15)$ in $\Phi(210)$ contains a $G_{2213}$ 
singlet which is even under D-parity \cite{cmp}. When this component acquires VEV, $SO(10) \to G_{2213}$ with unbroken left-right discrete symmetry.

Unlike the D-parity breaking case where the intermediate left-right gauge
group  has four different coupling constants, in the present case $G_{2213}$
has only three gauge couplings, $g_{2L}=g_{2R}, g_{3C}$, and
$g_{BL}$  for $\mu \ge M_R$. 
In the second step, the right-haded triplet component in 
$\overline {126}$
acquires VEV to break $G_{2213}\to G_{213}$ while generating heavy right-handed
Majorana neutrino mass. In the process of spontaneous electro-weak symmetry
breaking driven by weak bi-doublet in ${\bf {10}}$, small left-handed neutrino masses are 
generated through Type I and Type II see-saw mechanisms.

To discuss gauge coupling unification we use the following three RGEs 
from $M_Z$ to $M_U$ \cite{gqw,langacker,mkp1},

\ba
{1\over\alpha_Y(M_Z)}&=&{1\over\alpha_G}+{a_Y\over 2\pi}\ln{M_R\over M_Z}
+\frac{1}{10\pi}\left(3a'_{2L} + 2a'_{BL}\right)\ln{M_U\over M_R}
 ,\nonumber\\ 
{1\over\alpha_{2L}(M_Z)}&=&{1\over\alpha_G}+{a_{2L}\over 2\pi}\ln{M_R\over M_Z}
+ {a'_{2L}\over 2\pi}\ln{M_U\over M_R},                                          \nonumber\\ 
{1\over\alpha_{3C}(M_Z)}&=&{1\over\alpha_G}+{a_{3C}\over 2\pi}\ln{M_R\over M_Z}
+{a'_{3C}\over 2\pi}\ln{M_U\over M_R}.\label{rges}                       
\ea 
\noindent
where $\alpha_G$ is the GUT fine-structure constant and the beta function
 coefficients $a_i$ and $a'_i$ are determined by the particle  spectrum 
 in the ranges  from $M_Z$ to $M_R$, and from  $M_R$ to $M_U$,
 respectively.  
 Adopting the standard procedure we  obtain the following two equations
~\cite{mkp1}

\ba
L_{\theta} &\equiv&\frac{2\pi}{\alpha(M_Z)}\left(1- \frac{8 ~Sin^2\theta_W(M_Z)}{3}
\right)~=~ A\ln{M_U\over M_Z}+ B\ln{M_R\over M_Z},\label{eqs}  \\
L_S &\equiv&\frac{2\pi}{\alpha(M_Z)}\left(1- \frac{8\alpha(M_Z)}{3\alpha_{3C}(M_Z)}
\right) ~=~ A'\ln{M_U\over M_Z}+ B'\ln{M_R\over M_Z}. \label{eqc}
\ea
 where 
\ba
A &=& {2\over 3}(a'_{BL} - a'_{2L}), ~B = {5\over 3}(a_Y
-a_{2L})- {2\over 3}(a'_{BL} - a'_{2L}), \nonumber \\
A'&=& 2a'_{2L} +{2\over 3}a'_{BL}-{8\over 3}a'_{3C}, ~B'= {5\over 3}a_Y
+a_{2L} - {8\over 3}a_{3C} -  (2a'_{2L} +{2\over 3}a'_{BL}-{8\over 3}a'_{3C}).
\label{eqab}
\ea

From eq.(\ref{eqs}) and eq.(\ref{eqc}), the analytic expressions for $M_U$ and 
$M_R$ immediately follow,

\ba
\ln{M_U\over M_Z} &=& \frac{1}{(AB'-A'B)}\left (B'L_{\theta}-B L_S\right ),
 \label{eqmu}\\
\ln{M_R\over M_Z} &=&  \frac{1}{(AB'-A'B)}\left(AL_S - A'L_{\theta}\right). 
\label{eqmr}
\ea

Using PDG values, $\alpha(M_Z) = 127.9, ~Sin^2\theta_W(M_Z)= 
0.2312$ , and
$\alpha_{3C}(M_Z) = 0.1187$ \cite{pdg}, we obtain  

\be
L_S  = 662.736, ~~L_{\theta} = 308.305     .\label{data} 
\ee

Ignoring the contributions from additional lighter scalar multiplets emerging from mass spectra
analysis which have been  discussed later,
 the minimal Higgs  content necessary to break 
SUSY $SO(10)$ through $G_{2213}$ to the low energy group
and the associated beta function coefficients are,

\par\noindent{\bf {$\mu = M_Z$ - $M_R$:}}\\

\ba
H^u(2, 1,1)\op H^d(2, -1,1)  \subset G_{213},\\
a_Y = 33/5, a_{2L} = 1, a_{3C} = -3,  \label{eqz}
\ea

\par\noindent{\bf {$\mu =  M_R - M_U$:}} \\
\ba
H^{\phi}(2,2,0,1), \Dl(3,1, -2, 1)\op \Dr(1,3 , -2, 1)\op 
\ol {\Dl}(3,1, 2,1) \op  \ol {\Dr}(3,1, 2, 1),\nonumber \\
\ea
under $G_{2213}$ with 

\ba
a'_{BL} &=& 24, ~a'_{2L} =a'_{2R} =5, \nonumber\\ 
a'_{3C} &=& a_{3C} =-3. \label{eqr}
\ea  

Thus, the one-loop coefficients give,

\be
A=38/3,~B=-10/3, ~A'=34, ~B'=-14, ~AB'-A'B=-64 ,\label{eqAB}
\ee

leading to the solutions \cite{majee}, 
\be
M_R \simeq 10^{16} ~~GeV, ~M_U = 2\times 10^{16} ~~GeV. \nonumber
\ee
 The mass spectra analysis in the  minimal SUSY $SO(10)$ with $210\op
 126\op\ol {126} \op 10$ predicts that there are  additional scalar components 
of $210$ having intermediate scale mass with  the following $G_{2213}$
quantum numbers \cite{bajc2},
\be 
(3,1,-2/3,3),(3,1,2/3,\ol 3), (1,3,-2/3,3),(1, 3,2/3, \ol 3) \label{eqadd}\\   
\ee
 Not only these states prevent any value of intermediate scale below $M_U$,
as explicitly noted in Ref.~\cite{bajc2}, but, as noted in Ref. 
\cite{majee}, their presence at scales substantially lower than $2\times 10^{16}$ GeV
also spoils perturbative renormalization of  gauge couplings.

 It has been  pointed out that $G_{2213}$ intermediate scale
can still be obtained in non-minimal $SO(10)$ by threshold
effects  or by the presence of  non-renormalizable dim.5 operators in the
$SO(10)$ Lagrangian \cite{shafi,majee}. These might arise if, in addition to 
$\bf {210}$, the theory
contains a Higgs representation $\bf {54}$. The implementation of the gauge coupling
unification has been found to be  possible by threshold corrections or 
by gravitational
corrections in non-renormalizable SUSY SO(10) only if the additional light scalars given in 
eq. (\ref{eqadd}) are made superheavy which could be realized  due to the added
presence of $\bf {54}$ \cite{majee}.

However, the more attractive and  popular unification scheme being through  manifestly one-loop 
evolution of  gauge couplings in a  renormalizable grand unified theory
where coupling constants from lower scale evolve to converge to the unification coupling at the GUT scale,
in the next two sections we show how such a  unification is achieved 
 when the flavor symmetry $S_4$ is combined with $SO(10)$ as well as  
the R-parity and parity conserving supersymmetric left-right gauge theory $G_{2213}$ at the intermediate scale.

\section{III. Intermediate Left-Right Gauge Symmetry with  
  $S_4$  Flavor Symmetry}

In this section we show that in the presence of flavor symmetry 
and
left-right gauge symmetry, the extrapolation of standard model gauge couplings
through $G_{2213}\times S_4$   intermediate  symmetry naturally 
leads to successful unification of gauge couplings at $2\times 10^{16}$ GeV.
In the next section we show how the minimal particle content necessary for this
intermediate symmetry follows from mass spectrum analysis of supersymmetric
$\gf$.

  It is well known that  the minimal particle content of the standard 
model (SM) alone in non-supersymmetric theory  does not 
 allow its gauge couplings to unify at any higher scale. However, when the 
SM emerges from nonSUSY left-right symmetric $G_{2213}$ at the
intermediate scale, ~profound unification of gauge coplings occurs at
the GUT scale \cite{lmpr}.
Since the non-SUSY theories are well known for their generic gauge hierarchy problem,  when supersymmetry is combined with the SM to solve this problem, 
the enlarged particle spectrum naturally infused into the MSSM, ~achieves 
gauge coupling unification at the SUSY GUT scale without any intermediate scale. ~On the other hand, if an intermediate symmetry with extended gauge group like $G_{2213}$ is introduced, manifest  one-loop gauge coupling unification is 
spoiled in SUSY GUTs.
  
Earlier, without ascribing any connection with flavor symmetry,
 several authors have noted  that when the particle spectrum
at lower scales is further extended   beyond the minimal spectrum, 
manifest one-loop gauge coupling unification occurs in the presence of $G_{2213}$ intermediate symmetry with or without parity or R-parity
 \cite{ majee, lm}.

 These observations lead  us to suggest that the present failure  to achieve 
manifest one-loop gauge coupling unification with  intermediate left-right 
gauge symmetry  may be hinting at its extension to include a family symmetry with corresponding extesion in the particle spectrum.
 We find that this new symmetry to be appended to $G_{2213}$ and $SO(10)$
could be the well known flavor symmetry $S_4$ \cite{lee, hagedorn, mpr}.

 This conclusion can be inferred by also looking into the structure of 
the RGEs 
in eq.(\ref{eqs}) and eq.(\ref{eqc}). It is clear that if the coefficients
$B$ and $B'$ are negligible compared to $A$ and $A'$, then the solutions for
the mass scales would be insensitive to the values of $M_R$. Then values of $M_R$ substantially lower than the SUSY GUT scale could be tolerated by the RG constraints. ~We find that this possibility can be realised within ${\gf}$.

 In the  enlarged particle spectrum, 
along with the two sets of Higgs  triplets of the minimal scenario given in 
eq.(\ref{eqr}),    
the presence of $S_4$ symmetry needs six bi-doublets instead of only one
 \cite{hagedorn}. In additon,  other scalar multiplets 
belonging to $\bf {1}$ , $\bf {2}$ , or $\bf {3}$ of $S_4$ having 
nontrivial transformation property under $G_{2213}$ are also found to  be essential.
These latter Higgs particles turn out to be a triplet $\bf {3}$ of $S_4$ 
and each member of the trplet  transforms as a color octet. The fermions of three generations are taken as ${\bf 3}$' of $S_4$.

Thus, keeping the MSSM particle spectrum from $M_Z$ to $M_R$ unaltered, the 
enlarged  Higgs spectrum at the intermediate scale consistent with 
$G_{2213}\times S_4$ symmetry is:\\

\par\noindent {\bf {$\mu =  M_R - M_U$:}}\\

\ba
\Dl(3,1, -2, 1)\op \Dr(1,3 , -2, 1)\op 
\ol {\Dl}(3,1, 2,1) \op  \ol {\Dr}(3,1, 2, 1),\nonumber \\
6(2,2,0,1), 3(1,1,0,8),\label{eqps} 
\ea
where ${\bf 6=3+2+1}$, and ${\bf 3,2}$ and ${\bf 1}$ are triplet, doublet, and  singlet ,
respectively, under $S_4$. These Higgs scalars modify the beta-function 
coefficients to \\

\be
a'_{BL} = 24, ~a'_{2L} =a'_{2R} =10,  ~a'_{3C} = 6 .\label{eqfr}
\ee

Noting that the value of $a'_{BL} = 24 $  which is the same as in
eq.(\ref{eqr}), determines the perturbative constraint on the lowest allowed
value of $M_R$  \cite{majee}, we have with the above Higgs spectrum,  

\be
A = {28\over 3}, ~A' = 20 , ~B = ~B' = 0. \label{eqfab}
 \ee 
 
It is interesting to note that the particular combination of 
$G_{2213}\times S_4$ given in eq.(\ref{eqps}) leads to exactly vanishing values of $B$ and $B'$. 
The fact that now $B$ and $B'$ both vanish with such Higgs content ensures the possibility of an 
intermediate scale over wide range of values. 
But the perturbative lower bound on $M_R$ being determined by $a'_{BL}=24$
due to the appearance of a Landau pole,
now the popular one loop unification of gauge couplings is expected to 
 materialise for  values of  the intermediate scale 
satisfying this bound \cite{majee}.

\begin{center}
\begin{figure}[thb]
\psfig{figure=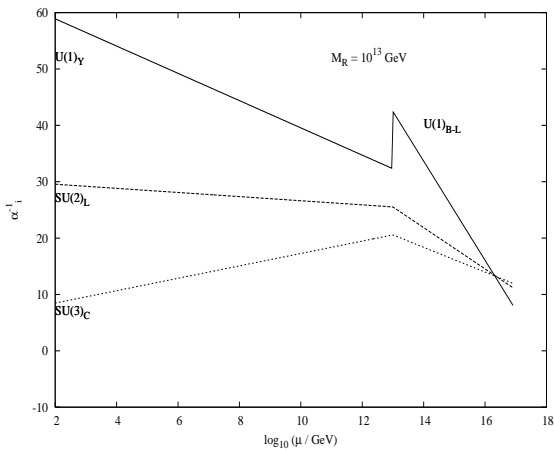,width=7.5cm,height=6.0cm,angle=270}
\caption{Evolution of gauge couplings showing variation of inverse fine-structure constants $\alpha_i^{-1}(\mu)$ as a function of the mass scale $\mu$
with $G_{2213}\times S_4$ 
intermediate
gauge symmetry breaking at $M_R=10^{13}$ GeV  in SUSY $\gf$ model. The top solid line represents $\alpha_Y^{-1}(\mu)$ for $\mu = M_Z - M_R$ and 
$\alpha_{B-L}^{-1}(\mu)$
for $\mu = M_R - M_U$. The middle and the bottom lines represent 
$\alpha_{2L}^{-1}(\mu)$ and $\alpha_{3C}^{-1}(\mu)$, respectively,  throughout the range of $\mu$. } 
\label{fg1}
\end{figure}
\end{center}

\begin{center}
\begin{figure}[thb]
\psfig{figure=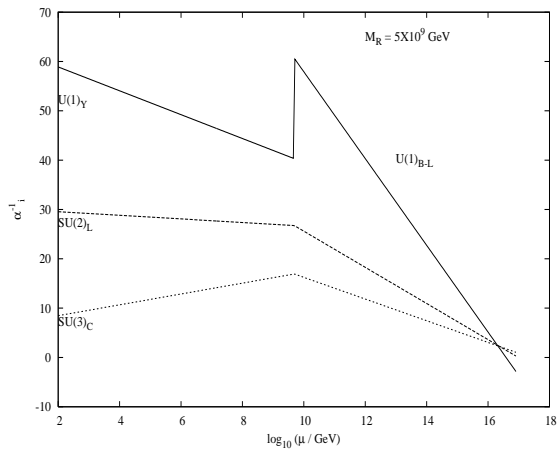,width=7.5cm,height=6.0cm,angle=270}
\caption{Same as Fig. \ref{fg1} but for $M_R=5 \times 10^9$ GeV.} 
\label{fg2}
\end{figure}
\end{center}

As eq.(\ref{eqmu}) and eq.(\ref{eqmr})
are no longer valid with $~B = ~B' = 0$ , we solve for the mass scales numerically  using eqs.(2), (10), and (17). We find all
values of the left-right symmetry breaking  scale $M_R$ are permitted 
over a wide range,
\be
5\times 10^{9} ~{\rm GeV} \le  ~M_{\rm R}~  \le ~10^{16} ~{\rm GeV}.  
\label{bound}
\ee 
but having almost the same value of unification scale $M_U = 2\times 10^{16}$ GeV 
for all solutions. Two examples of such solutions for
$M_R =  10^{13}$ GeV and $M_R = 5\times 10^9$ GeV are
  shown in Fig. \ref{fg1} and Fig. \ref{fg2}, respectively.
where manifest one-loop  unification with the three gauge couplings of
$G_{2213}$ converging at the GUT scale, $M_U = 2\times 10^{16}$ GeV, is evident.

For $M_R \simeq 10^{15}$ GeV , the  GUT fine structure constant has the value 
$\alpha_G \simeq 1/24.5$ which increases as $M_R$ approaches lower values, reaching $\alpha_G \simeq 1/12$ and $\alpha_G \simeq 1/2 $ at $M_R=10^{13}$ GeV and
 $M_R=5\times 10^9$ GeV, respectively. This phenomenon is  due to the 
appearance of Landau pole near $M_U = 2\times 10^{16}$ GeV in the gauge 
coupling of $U(1)_{B-L}$  when $M_R \le 10^9$ GeV. Now that all the three gauge couplings of $G_{2213}$
are unified at the GUT scale, $\alpha_G$ would approach $\infty$ for the same value of $M_R \le 10^{9}$ GeV. The basic reason is that $a'_{B-L} = 24$  has 
remained the same as the minimal model inspite of new contributions from 
 $G_{2213}\times S_4$ multiplets.

The $SU(2)_R$ gauge coupling at $\mu \simeq M_R$ is nearly $g_L = g_R = g
\simeq 0.7 $. With the $SU(2)_R$ gauge boson mass scale $M_R \simeq gv_R$,    
the vacuum expectation value of the Higgs triplet field is also allowed in the 
similar range with $<\overline {\Delta_R}^0> = v_R \simeq 5\times 10^9$ GeV to
$10^{16}$ GeV. Ignoring the constraint from the neutrino oscillation data
which will be discussed in Sec.6 ,
the right-handed neutrino mass $M_N=f_0v_R$ for Majorana 
coupling $f_0 \simeq 1$ is then allowed to vary over similar range without
the necessity of any  tuning of $f_0$.


In the next section we show how the minimal particle content needed for 
gauge coupling and flavor unification through $G_{2213}\times S_4$  can be
easily embedded in $\gf$. We further show how the F-term flatness condition
leads to the desired minimal particle spectrum of the model below the GUT
scale  while keeping all other 
components of the $\gf$ representations superheavy.    
 
\section{IV. Symmetry Breaking  of $SO(10)\times S_4$ }

We consider $S_4$ flavor symmetry for three fermion generations and supersymmetric grand 
unification of three forces of nature  through  $SO(10)
\times S_4$ \cite{lee, hagedorn}. Instead of starting from flavor symmetric
 standard model  without SUSY discussed in Ref. \cite{hagedorn},  
we consider flavor symmetry starting at the intermediate scale through $G_{2213}\times S_4$. Such a model leading to MSSM at lower scales  originates from 
intermediate
  breaking of $SO(10)\times S_4$,

{\Large $SO(10)\times S_4 \mathop{{\rightarrow}^{210}_{M_U}} ~~G_{2213}\times S_4$}  
{\Large $\mathop{{\rightarrow}^{126+\ol {126}}_{M_R}}~~G_{213} 
\mathop{{\rightarrow}^{10}_{M_W}}~~U(1)_{em}\times SU(3)_C$}.\\

To achieve successful gauge coupling unification with the desired 
intermediate scale,
we need the minimal particle content same  as in the 
MSSM from $M_Z$ to $M_R$. In the presence of $G_{2213}\times S_4$,  we need
six bi-doublets each of which is embeddable in $H(10)_0, H(10)_{1,2}, H(10)_{3,4,5}$ 
of $SO(10)$ and these transform  as  singlet, doublet, and triplet,
respectively,  under $S_4$. The left and the
right-handed Higgs triplets needed to maintain supersymmetry and implement
spontaneous breaking of $G_{2213}$ at  scale $M_R$ are contained in 
$\Sigma_0(126)$ and $\ol{\Sigma_0}(\ol {126})$ 
 of $SO(10)$.
We note that, in addition to the minimal number of doublets and triplets, 
 the gauge coupling unification also needs three $SU(3)_C$ octets 
transforming as $(1, 1, 0, 8)$ under $G_{2213}$. 
Each of them is contained in the $G_{224}$ multiplet $(1,1,15)$ which, in turn,
is contained in the $A_i(45)~(i=1,2,3)$ of $SO(10)$ treated as a 
triplet ${\bf  3}$  of $S_4$.
All other Higgs particles are  to become 
superheavy with masses near the GUT scale for  successful gauge coupling 
unification at one-loop level and this can be achieved in the 
presence of $S(54)$ of $SO(10)$~\cite{goh3}. The Fermion and Higgs representations are given in Table \ref{tab1}.

\begin{table*}
\caption{Particle content of the model and their 
transformation properties under $S_4\times SO(10)$   }   
\begin{ruledtabular}
\begin{tabular}{lcccccccc}\hline
$\bf {Fermions}$&& $\bf {Higgs ~Bosons}$&&&\\\hline
${\bf \Psi_i, (i=1,2,3)}$&${\bf S}$&${\bf \Phi }$&
${\bf A_{1,2,3}}$&$\bf {\overline {\Sigma}_0 \oplus {\Sigma}_0 }$&
$\bf  {H_0}$&$\bf {H_{1,2}}$&$\bf {H_{3,4,5}}$\\
\hline\\
${\bf {3^{\prime}}}\times {\bf {16}}$&${\bf {1\times 54}}$&${\bf {1}}\times {\bf {210}}$&${\bf { 3\times 45}}
$ &
${\bf {1}}\times {\bf \overline {126}\oplus 126}$&${\bf {1}}\times {\bf {10}}$&${\bf {2}}
\times {\bf {10}}$&${\bf {3}}\times {\bf {10}}$\\
\end{tabular}
\end{ruledtabular}
\label{tab1}
\end{table*}

\par
In order to realize  such a spectrum by actual potential 
minimization in the presence of supersymmetry,  we break 
$SO(10)$  by giving GUT-scale vacuum expectation values to the two
D-parity conserving singlets of $\Phi(210)$ and $S(54)$.~We follow the mass spectrum analysis technique  for supersymmetric $SO(10)$ grand unification
\cite{fuku1,  bajc2,  fuku2}.  
We assign nearly equal vacuum expectation values to both the singlets 
with $\langle S \rangle  \simeq \langle \Phi  \rangle$ such that, effectively, 
the GUT symmetry breaking to $G_{2213}$ appears as one step process.

Although for the sake of gauge coupling unification alone, it is
possible to treat the effective gauge symmetry below $M_R$ as
MSSM$\times S_4$ with only one weak bi-doublet having mass near the
electro-weak scale, for accommodating all fermion masses and mixings
in this model as dicussed in the next section, it is necessary to
break the flavor symmetry at the intermediate scale.

 We assume that  $G_{2213}$ gauge symmetry is broken 
at the scale $M_R$
 due to vacuum
expectation value of the standard model singlet contained in the right
handed triplet  $\overline {\Delta}_R(1,3, 2 ,1)$
in  $\overline{126}$. After the $G_{2213}\times S_4$ breaking, only two
MSSM Higgs doublets are taken  to remain light.

The superpotential near the GUT scale can be written as,
\ba
W_{H} &=& \frac{1}{2} m_{\Phi}\Phi^2 +  \frac{1}{2} m_{S} S^2 +  \frac{1}{2}
m_{A}{\sum}_i A_i^2 + m_{\Sigma}\Sigma_0 \ol {\Sigma}_0 +  \frac{1}{2}m_{H_0}H^2_0
 \nonumber \\
&& +\frac{1}{2}m_{H_D}{H_D}^2 + \frac{1}{2}m_{H_T} {H_T}^2
+\lambda_0\Phi^3 + \lambda_1\Phi \Sigma_0 {\ol {\Sigma}_0}+ (\lambda_2\Sigma_0
 +\lambda_3\ol {\Sigma}_0)H_0\Phi 
 + \lambda_4 {\sum}_i A_i^2\Phi\nonumber \\ 
&& + S(\lambda_5S^2 +\lambda_6{\sum}_i A_i^2 +\lambda_7\Phi^2+\lambda_8
\Sigma^2_0 + \lambda_{9} \ol{\Sigma}_0^2 + \lambda_{10}H_0^2 + \lambda_{11}H_D^2
 + \lambda_{12}H_T^2). \label{sup}
\ea

Using vacuum expectation values $<S>= v_S, <\Phi>= v_{\Phi},
<\dr> = \sg , <\drb> = \sgb$,
the vanishing F-terms yield \cite{fuku2},

\ba
m_{\Phi}v_{\Phi} + \frac {\l_0v_{\Phi}^2}{3\sqrt 2} + \frac{\l_1\sg\sgb}{10\sqrt 2}
-\frac{2\l_7 v_{\Phi}v_S}{\sqrt {15}} &=& 0, \label{eqf1} \\
m_Sv_S + \frac{\sqrt 3 \l_5 v_S^2}{2\sqrt 5} -\frac {\l_7v_{\Phi}^2}{\sqrt {15}}
&=& 0, \label{eqf2} \\
\left [m_{\Sigma} + \frac{\l_1v_{\Phi}}{10\sqrt 2} \right ]\sg &=& 0.\label{eqf3}
\ea 
 
Due to the vanishing D-term, $\sg = \sgb\equiv v_R$ and the corresponding F-terms for 
$\sg$ or $\sgb$ yield the same equation as  eq.(\ref{eqf3}).
In the desired hierarchial case, both $\sg$  and $\sgb$ are much 
smaller compared to $<S>, <\Phi>$ leading to the relation between the
GUT-scale VEVs and $m_{\Phi}$, 

\be
m_{\Phi} + \frac {\l_0v_{\Phi}}{3\sqrt 2}  
-\frac{2\l_7 v_S}{\sqrt {15}} = 0. \label{eqf4} \\
\ee

Using $v_{\Phi}$ from 
eq.(\ref{eqf4}) in eq.(\ref{eqf2}) 
gives a quadratic equation for $v_S$,

\be
p v_S^2 + q v_S - r = 0, \label{qdr}
\ee
where
\ba
p &=& \frac{\sqrt 3 \l_5}{2\sqrt 5}-\frac{24 \l_7^3}{5\sqrt {15}\l_0^2},
\nonumber\\
q &=& m_S + \frac{24 \l_7^2}{5\l_0^2} m_{\Phi},\nonumber\\
r &=& \frac{18 \l_7}{\sqrt {15}\l_0^2}m_{\Phi}^2. \label{pqr}
\ea  

In the next section we discuss the emergence of mass spectra necessary to 
 keep only the desired minimal number of  Higgs particles  light while making 
others superheavy.

\section{V. Light and Heavy Particle States from Mass Spectra}  
 
In this section we discuss the emerging mass spectra from the spontaneously
broken flavor symmetric GUT while making provisions for would be Goldstone bosons and
the light scalars necessary for gauge coupling unification. In contrast
to the minimal model without flavor symmetry where unwanted 
light scalar degrees of freedom are found to spoil gauge coupling unification
 \cite{bajc2}, 
in the present case, 
due to the presence of the scalar multiplet ${\bf {54}}$ in the Higgs superpotential, it is
possible to lift those masses to the GUT scale.

\subsection{A. Goldstone Bosons} 

In the process of spontaneous symmetry breaking of $SO(10) \to G_{2213}$
through $<S>$  and 
$<\Phi>$,
$30$ gauge bosons would aquire GUT-scale mass by absorbing the corresponding 
mass-less scalars. Under $G_{2213}$ these superheavy gauge bosons have the 
quantum numbers $(2,2,2/3,3)+(2,2,-2/3, \ol 3)+(1,1, 2/3,3)+ 
(1,1, -2/3,\ol 3)$. Whereas the first two sets of states are contained 
in both $(2,2,6) \subset G_{224}\subset 54$  and $(2,2,\ol {10}) \subset
G_{224}\subset 210$ of $SO(10)$, the next two sets of states are contained
in $(1, 1, 15) \subset G_{224}\subset 210$ or $45$ of $SO(10)$.  
Using the superpotential in eq.(\ref{sup}) and the vacuum expectation values,
we show how the desired Goldstone bosons are obtained.\\

\par\noindent{\bf{A.1. $(1, 1, 3, 2/3)+ (c. c)$ as Goldstone Bosons}}\\

Noting that $\sg = \sgb\equiv v_R << v_S \sim v_{\Phi} \sim M_U$, it turns out that 
these unmixed states in the leading approximation have masses,
\be
m_{G1}  = m_{\Phi} + \frac {\l_0v_{\Phi}}{3\sqrt 2}  
-\frac{2\l_7 v_S}{\sqrt {15}}. \label{g1} \\     
\ee

Using eq.(\ref{eqf4}) it is immediately recognised that these are 
naturally the light pseudo  Golstone bosons to be absorbed by the mass-less 
vector bosons to make them superheavy.
 It can be easily checked that other unmixed states having the same quantum
 nubers in $A_i(i=1,2,3)$ have degenerate superheavy masses,
 
\be
m_A + \frac{\sqrt 2 \l_4 v_{\Phi}}{\sqrt 3}- \frac{2\l_6 v_S}{\sqrt {15}}        ,   \label{ma1}
\ee    
 
which are naturally  near the GUT scale.\\

\par\noindent{\bf{A.2. (2, 2, 3, 1/3)+ (c. c) as Goldstone Bosons}}\\

Using the basis $ {\bf [ (A_1)_{(2,2,6)}^{(2,2,1/3,3)}, 
(A_2)_{(2,2,6)}^{(2,2,1/3,3)},(A_3)_{(2,2,6)}^{(2,2,1/3,3)},
S_{(2,2,6)}^{(2,2,1/3,3)}, \Phi_{(2,2, 6)}^{(2,2,1/3,3)},
\Phi_{(2,2, 10)}^{(2,2,1/3,3)}]}$ ~\cite{fuku2}
where the superscripts(subscripts) refer to gauge quantum numbers of the Higgs
multiplets under $G_{2213}(G_{224})$ , there are three unmixed pairs of states 
in 
$A_1$ and $A_2$ and $A_3$ with degenerate superheavy masses,

\be
m_A + \frac{\l_6}{2\sqrt {15}}v_S. \label{ma2}
\ee
  
The fourth unmixed pair  is that of $\Phi_{(2,2, 6)}^{(2,2,1/3,3)}$ with
superheavy mass,
\be
m_{\Phi} + \frac{7\l_7}{4\sqrt {15}}v_S. \label{mphi1}
\ee
The remaining two pairs of states,  $S_{(2,2,6)}^{(2,2,1/3,3)}\oplus (c.c.)$ and 
$\Phi_{(2,2, 10)}^{(2,2,1/3,3)}\oplus (c.c)$, mix through the mass matrix,

\par\noindent
\be M_1=\left[\br{cc}
m_S+\frac{\sqrt 3 \l_5v_S}{2\sqrt 5}&\frac{\l_7v_{\Phi}}{2\sqrt 3}\\
\frac{\l_7v_{\Phi}}{2\sqrt 3}&m_{\Phi}+\frac{\l_0v_{\Phi}}{3\sqrt 2}-\frac{\sqrt 3
\l_7v_S}{4\sqrt 5}\\
\er\right]. \label{mphis1}\ee
It is clear that one linear combination of these two pairs  can be made
mass-less by tuning the parameter $\l_5$ such that it supplies the remaining 
Goldstone modes. The other orthogonal combination acquires mass near the GUT scale.

\subsection{\bf B.  Light Scalrs from Mass Spectra for Gauge Coupling Unification}
 
In Sec.3, the unification of gauge couplings with $G_{2213}\times S_4$
intermediate gauge symmetry 
has been shown to require the usual left- and  the right-handed triplets
that are contained in $126\op \ol {126}$, six bi-doublets  contained in six 
$10$-plets, and a set of three color octets transforming as $(1,1,0,8)$
under $G_{2213}$. These octets are contained in the $G_{224}$ submultiplet
$(1,1,15)$ of $A_i(45)$ or $\Phi(210)$ of $SO(10)$. In addition, 
the $G_{224}$ submultiplet $(1,1,20')$ in $S(54)$ also contains the 
octet component. But we will find it covenient to obtain these three
octets from a triplet of $45_i(i=1,2,3)\subset SO(10)$.

At first it is to be noted that the triplets 
 $\Delta_L(3,1, -2 ,1), {\Delta}_R(1,3,-2 ,1)$ and their conjugates contained  
$126\op\ol{126}$   
acquire degenerate masses,
\be  
M_R = m_{\Sigma} + \frac{\l_1v_{\Phi}}{10\sqrt 2}. \label{mdel1}
\ee

The condition $M_R << M_U $  can be ensured by tuning $\l_1$.
At first the six bi-doublets from the six $10$-plets are treated to have masses near $M_R$
in  the usual fashion by some
 doublet triplet splitting mechanism or by tuning the parameters $\l_{10}, 
\l_{11}$, and  $ \l_{12}$ while the weak bi-doublets in   $126\op \ol
{126}$  and $210$  are kept heavy at the GUT scale. In the next
section we show how five linear combinations of these bi-doublets can
be treated to have masses at the $M_R$-scale while keeping the mass of
the remaining linear combination at the electro-weak scale, thus
supplying the pair of two MSSM doublets $(H^u, H^d)$.

Choosing the basis {\bf {$(A_i)_{1,1,15}^{1,1,0,8}, S_{1,1,20'}^{1,1,0,8},
\Phi_{1,1,15}^{1,1,0,8}$}}, we find that there are three  unmixed 
states in $A_i$ with masses,

\be
m_A -\frac{\sqrt 2 \l_4 v_{\Phi}}{3} - \frac{2 \l_6 v_S}{\sqrt {15}}. 
\label{ma3} 
\ee
 Clearly the advatage of $A_i$ being the members of  ${\bf {3 \subset S_4}}$ 
is that the tuning of the single parameter $\l_4$ makes all the three 
 octets light having masses near the $M_R$ scale which is essential for gauge coupling unification. 
It is found that the other two states mix through the mass matrix,

\par\noindent
\be M_2=\left[\br{cc}
m_S-\frac{2\sqrt 3 \l_5v_S}{\sqrt 5}&-\frac{\l_7 v_{\Phi}}{\sqrt 6}\\
-\frac{\l_7 v_{\Phi}}{\sqrt 6}&m_{\Phi}-\frac{\l_0 v_{\Phi}}{3\sqrt 2}-\frac{2
\l_7v_S}{\sqrt {15}}\\
\er\right]. \label{mphis2}\ee
The eigen values emerging from eq. (\ref{mphis2}) are  at the GUT scale and we do not adopt any
further fine-tuning.

We have verified that all the components  from $G_{224}$ multiplets,$(1,3,15)\op (3,1,15)\subset \bf {210}$ 
 acquire masses near the GUT scale due to the
presence of ${\bf {54}}$ in the model and there are no other lighter states 
which are likely to disrupt gauge coupling unification.

In summary the theory has enough parameter space for the successful
implemetation of the model  with $G_{2213}\times S_4$ intermediate 
symmetry and gauge coupling unification at the GUT scale via $\gf$.

\section{VI. Fermion Masses and Mixings}

In this section we address the question of fermion masses and mixings and make 
predictions in the neutrino sector. 
For this purpopose we assume type-I seesaw dominance and also include a pair of $\bf{{126}_{1,2}\oplus \ol {126}_{1,2} \equiv 
{\Sigma}_{1,2}\oplus\ol {\Sigma}_{1,2}}$ as members of an $S_4$
doublet but with all their components having GUT 
scale masses \cite{lee1}. Even though all the weak bi-doublets
    $\bf {\overline {\Delta}^{\phi}_i  (i=0,1,2)}$
 contained in the $G_{224}$ submultiplets $\bf {(2,2,15)}$ of
$\ol {\Sigma}_i, (i=0,1,2)$ have GUT scale masses,  it is necessary to clarify 
 how   their  
induced VEVs contribute to the Dirac masses of all fermions in a manner 
analogous to minimal SUSY  $SO(10)$ \cite{babu1}.

\subsection{\bf A. Light Weak Bi-doublets  
and Vacuum Expectation Values}

In this subsection we clarify  how by keeping only the desired
minimal number of particles below the GUT scale, VEV of different weak
bi-doublets are made available to generate fermion masses withot
disrupting gauge coupling unification.  
   
The added $S_4-$
doublet fields $\bf{\Sigma_D\oplus \ol {\Sigma}_D}$ will have a new 
contributions to the superpotential which include,

\be
 W'_H=     m_{\Sigma'}\Sigma^T_D \ol {\Sigma_D} + 
 + \lambda'_1\Phi \Sigma^T_D {\ol {\Sigma}_D}+ (\lambda'_2\Sigma^T_D +
\lambda'_3\ol {\Sigma^T_D})H_D\Phi........ \nonumber
\ee   
where ellipses denote couplings to other Higgs representations and subscript D 
stands to indicate a $S_4-$doublet. Since  $m_{\Sigma'} \simeq M_U$
and we need no fine-tuning of parameters $\lambda'_{1,2,3}$ to
maintain gauge hierarchy, it immediately follows that all the
components of the $S_4-$ doublet pair (including the weak bi-doublets
and triplets) acquire masses at the GUT scale  and they do not upset
gauge coupling unification.
Denoting the full superpotential as $\bf {W=W_H+W'_H}$, the $\bf {F-}$term due to $\bf{\Phi}$
now contributes the following terms to the scalar potential,
\ba
V&=&(\lambda_2\Sigma_0 +
\lambda_3\ol {\Sigma_0})H_0\Sigma_0\ol {\Sigma}_0 \nonumber \\  
&&+(\lambda'_2\Sigma^T_D +
\lambda'_3\ol {\Sigma^T_D})H_D\Sigma_0\ol {\Sigma}_0. \label{induced}
\ea
 This has the implication that whenever the RH-triplets in 
$\bf {\Sigma_0\oplus\ol {\Sigma}_0}$ and the weak doublet in  $\bf
{S_4}$-singlet $\bf{H_0\subset {10}_0}$ aquire VEVs  $v_R$ and
$\alpha_0 \equiv y_0<H_0>$, respectively, where the latter is
approximately of the order of the weak scale, the weak doublets 
in $\bf {\ol {\Sigma}_0}$ gets an induced VEV. On the otherhand, 
in addition to $v_R$, 
 the  VEVs of the $\bf{S_4-}$ doublet components $\bf{H_{1,2} \subset {10}_{1,2}}$
generate  induced VEVs in the weak doublets in $\ol {\Sigma}_{1,2}$. The order of magnitudes of all the three induced VEVs can now be expressed as,
\ba
\langle \Delta^{\phi}_i\rangle = {v_R^2\over M_U^2}{\alpha_i\over y_i}, 
(i=0,1,2).\label{qivev}
\ea    
where, as defined subsequently in this section,
$\alpha_i$ stands for the product of $\rm {i}^{\rm {th}}$  VEV and the
respective Yukawa coupling $y_i$.

One major difference from the minimal SUSY $SO(10)$ is that in
addition to  the VEVs 
of up- and down type doublets of $H_0$, the VEVs of nontrivial $\bf {S_4}-$
representations also enter into the RHS of eq.(\ref{qivev}). 

The next point that needs explanation is how only the six bi-doublets
lighter than $M_U$ acquire nearly electroweak-scale VEVs while five of
them have masses near $M_R$ scale and only the remaining bi-double  has
mass near the electro-weak scale to supply the up-type and the
down-type MSSM doublets ($H^u, H^d$). 

In order to achieve this objective we introduce two $SO(10)$-singlet
scalar fields, $\eta_S$ and $\eta'_S$ which transform as doublet and
triplet, respectively, under $S_4$. These will make additional
contribution to the superpotential,

\be
 W''_H= \lambda_S\eta_S H_0H_D +  \lambda'_S\eta'_S H_0H_T +.... \nonumber
\ee
   
We assign order $M_R$ scale VEVs to $\eta_S$ and $\eta'_S$ to break
the $S_4$ symmetry at the intermediate scale and generate $H_0-H_D$
and  $H_0-H_T$ mixings. Then using bi-unitary transformation on the
six doublet fielda to diagonalize the bi-doublet mass matrix at the 
intermediate scale, we treat only one  linear combinations of the weak
bi-doublets to have mass at the electro-weak scale  while the
remaining five linear combinations are treated to acquire mass at the
intermediate scale. This would involve only one fine-tuning of the new
parameters of the superpotential.
Once the lightest linear combination of all the six bi-doublets
is constructed in this manner to supply the MSSM Higgs 
doublets ($H^u, H^d$), the electroweak VEVs of the latter imply 
VEVs of approximately the same order for all the bi-doublet
components in $H_i(i=0,1,...,5)$. Then the induced VEVs of weak
bi-doublets at the GUT-scale contained in  
$(\Sigma \oplus \ol {\Sigma})_{0,1,2}$, already 
discussed in  eq.(\ref{qivev})
follow in a straight-forward manner.

To have a rough idea of the order of the VEVs involved, using
eq.(\ref{qivev})
and  taking the  respective Yukawa couplings in the range  
$y_i\simeq 0.01 - 1.0$ and $\alpha_i \simeq 100$ GeV, $v_R= 10^{13}$ GeV to
 $10^{14}$ GeV, $M_U= 2 \times 10^{16}$ GeV, we get $\langle \Delta^{\phi}_i\rangle
 = 10$ MeV to $10$ GeV. Our numerical analysis approximately agrees
with these results. 
   
\subsection{\bf B. Fermion Masses from $S_4$ Flavor Symmetry}

Investigation on fermion masses and mixings using an $\gf$ model but without any intermediate gauge symmetry has been carried out in Ref. \cite{cai} where
RG-extrapolated values of charged fermion masses at the GUT scale have been used to fix certain
model parameters and make predictions in the neutrino sector.
Although $v_R=10^{13}-10^{14}$ GeV  has been assumed with a view to obtain 
the right-handed neutrino mass $M_N=f_0v_R =10^{13}-10^{14}$ GeV for $f_0 \simeq 1$, we note that it is difficult to 
visualise any such value of $v_R$ substantially lower than the GUT-scale in a
single-step breaking scenario; hence the 
desired value of $M_R$ is not obtainable without adjusting the value of the
Majorana coupling to  $f_0 \approx 0.001 - 0.01$. Also we note that the right choice for the input values of charged fermion masses is desirable to be at the intermediate seesaw scale rather than the GUT scale. We carry out  investigations  utilising the RG-extrapolated values at 
$M_R \approx v_R \approx 10^{13}$ GeV in the present model  where no
adjustment of
$f_0$ is needed to obtain the desired see-saw scale. We  find that the experimental data on neutrino mass-squared differences and mixings in fact determine 
the seesaw scale to be 
 $M_N  \simeq 3.78\times 10^{13}$ GeV.  
 
Consistent with the $\gf $ symmetry  the 
 superpotential for fermion-Higgs Yukawa interaction is written as,
\ba
W^0_{Yuk}&=&(\Psi_1\Psi_1+\Psi_2\Psi_2+\Psi_3\Psi_3)(Y_0H_0+
f_0\overline{\Sigma_0})
+\frac{1}{\sqrt 2}( \Psi_2\Psi_2-\Psi_3\Psi_3)(y_1H_1+f_1\overline{\Sigma_1})\nonumber\\
&&+ \frac{1}{\sqrt 6}(
-2\Psi_1\Psi_1+\Psi_2\Psi_2+\Psi_3\Psi_3)(y_1H_2+f_1\overline{\Sigma}_2)
\nonumber\\
&&+y_3[(\Psi_2\Psi_3+\Psi_3\Psi_2)H_3+(\Psi_1\Psi_3+\Psi_3\Psi_1)H_4+
(\Psi_1\Psi_2+\Psi_2\Psi_1)H_5].\label{wy0}
\ea
\par\noindent
Following the standard notation with $Q(Q^C)$ and $L(L^C)$ for 
left(right)-handed quark
and lepton doublets in left-right symmetric gauge theory, and denoting 
$H_i^{\phi}$ and $\overline {\Delta}^{\phi}_i$ as the electroweak bi-doublets
in $H_i$ and $\overline {\Sigma}_i$, we now write the   Yukawa 
superpotential  consistent with $G_{2213}\times S_4$ just below the
GUT-symmetry breaking scale for $\mu \sim M_U$,
 before the electroweak
bi-doublets  $\overline {\Delta}^{\phi}_i$ decouple from the
superpotential,

\ba
W_{Yuk}&=&\Sigma_{k=1}^3[Q_k^T\tau_2(y_0 H^{\phi}_0+f_0\overline {\Delta}^{\phi}_0)
Q_k^C+L_k^T\tau_2(y_0H^{\phi}_0-3f_0\overline {\Delta}^{\phi}_0) L_k^C]
\nonumber\\
&&+\frac{1}{\sqrt 2}[ Q^T_2\tau_2(y_1 H^{\phi}_1+f_1\overline {\Delta}^{\phi}_1) Q_2^C- Q_3^T\tau_2(y_1 H^{\phi}_1+f_1\overline {\Delta}^{\phi}_2)Q_3^C\nonumber\\
&&+L_2^T\tau_2(y_1 H^{\phi}_1- 3f_1\overline {\Delta}^{\phi}_1)L_2^C -L_3^T\tau_2
(y_1 H^{\phi}_1-3f_1\overline {\Delta}^{\phi}_2)L_3^C]\nonumber\\
&&+
\frac{1}{\sqrt 6}[-2Q_1^T(y_1 H^{\phi}_2+f_1\overline {\Delta}^{\phi}_2)Q_1^C
-2 L^T_1\tau_2(y_1 H^{\phi}_2-3f_1\overline {\Delta}^{\phi}_2)L_1^C\nonumber\\
&&+Q_2^T\tau_2(y_1 H^{\phi}_2+f_1\overline {\Delta}^{\phi}_2)Q_2^C+L_2^T\tau_2
(y_1 H^{\phi}_2-3f_1\overline {\Delta}^{\phi}_2)L_2^C\nonumber\\
&&+Q_3^T\tau_2(y_1 H^{\phi}_2+f_1\overline {\Delta}^{\phi}_2)Q_3^C +L_3^T\tau_2
(y_1 H^{\phi}_2-3f_1\overline {\Delta}^{\phi}_2)L_3^C]\nonumber\\
&&+y_3[Q_2^T\tau_2H^{\phi}_3Q_3^C+Q_3^T\tau_2H^{\phi}_3Q_2^C+Q_1^T\tau_2H^
{\phi}_4Q_3^C
+Q_3^T\tau_2H^{\phi}_4Q_1^C+Q_1^T\tau_2H^{\phi}_5Q_2^C+Q_2^T\tau_2H^{\phi}_5Q_1^C\nonumber\\
&&+(Q \to L)]. \label{wy}
\ea

The up and down type electro-weak doublets in the six bi-doublets of $\bf {10}$'s $\subset SO(10)$
acquire VEVs $<H^u_i>=v^u_i$ and $<H^d_i>=v^d_i$($i=0,1,..,5$). The
electroweak submultiplets in  ${\overline {\Sigma}}_i$ also acquire
induced  VEVs
$<{\overline {\Delta}}^u_i>$ and $<{\overline {\Delta}}^d_i>, (i=0,1,2)$. 

Adding their 
contributions, the mass matrices of quarks and leptons have the well known 
forms, 
\be
M_u~=~M^{(10)}_u + M^{(126)}_u,~M_d~=~M^{(10)}_d + M^{(126)}_d,\nonumber\\
\ee
\be
M_l~=~M^{(10)}_d -3M^{(126)}_d,~M^D_{\nu}~=~M^{(10)}_u -3M^{(126)}_u,\nonumber\\\ee
\be
M_{\nu}~=~{M^D}^T_{\nu}M^D_{\nu}/M_N,\nonumber
\label{massdf}                       
\ee
\noindent 
where $M_N=f_0v_R$ is the degenerate right-handed neutrino mass and we have already noted in Sec.3 that we can  have $M_N\simeq M_R$ 
 substantially below the GUT-scale
 in this model without having the necessity to adjust the 
the Majorana coupling to be a small fraction of  unity .
The  component mass-matrix elements  in the above equations are defined as,
\par\noindent
\be M^{(10)}_u=\left[\br{ccc}
\alpha_0-2\alpha_2&\alpha_5&\alpha_4\\
\alpha_5&\alpha_0+\alpha_1+\alpha_2&\alpha_3\\ 
\alpha_4&\alpha_3&\alpha_0-\alpha_1+\alpha_2\\
\er\right], \label{mu10}\ee

\be M^{(10)}_d=\left[\br{ccc}
\beta_0-2\beta_2&\beta_5&\beta_4\\
\beta_5&\beta_0+\beta_1+\beta_2&\beta_3\\ 
\beta_4&\beta_3&\beta_0-\beta_1+\beta_2\\
\er\right], \label{md10}\ee

\be M^{(126)}_u=\left[\br{ccc}
\gamma_0-2\gamma_2&0&0\\
0&\gamma_0+\gamma_1+\gamma_2&0\\ 
0&0&\gamma_0-\gamma_1+\gamma_2\\
\er\right], \label{mu126}\ee

\be M^{(126)}_d=\left[\br{ccc}
\delta_0-2\delta_2&0&0\\
0&\delta_0+\delta_1+\delta_2&0\\ 
0&0&\delta_0-\delta_1+\delta_2\\
\er\right]. \label{md126}\ee
In these equations $\alpha_i \equiv y_i<H^u_i>, \beta_i \equiv y_i<H^d_i>,\gamma_i
\equiv f_i<\Delta^u_i>$ and $\delta_i \equiv f_i<\Delta^d_i>$
($i$ not summed). The choice of diagonal basis in the down quark sector which 
automatically also leads to the diagonal basis in the charged lepton sector,
enables to choose the six parameters, $\beta_i, \delta_i (i=0,1,2)$ to be real
and $\beta_3=\beta_4=\beta_5=0$. All other parameters are, in general, complex.
Analytically we express the six real parameters in terms of down-quark and charged lepton mass eigen-values at the see-saw scale ($\mu=M_R$),

\ba
\beta_0&=&\left[3(m^0_b+m^0_s+m^0_d)+m^0_{\tau}+m^0_{\mu}+m^0_e\right]/12,
\nonumber\\
\beta_1&=&\left[-3m^0_b+3m^0_s-m^0_{\tau}+m^0_{\mu}\right]/8,
\nonumber\\
\beta_2&=&\left[3m^0_b+3m^0_s-6m^0_d+m^0_{\tau}+m^0_{\mu}-2m^0_e\right]/24,
\nonumber\\
\delta_0&=&\left[m^0_b+m^0_s+m^0_d-(m^0_{\tau}+m^0_{\mu}+m^0_e)\right]/12,
\nonumber\\
\delta_1&=&\left[-m^0_b+m^0_s+m^0_{\tau}-m^0_{\mu}\right]/8,
\nonumber\\
\delta_2&=&\left[m^0_b+m^0_s-2m^0_d-m^0_{\tau}-m^0_{\mu}+2m^0_e\right]/24.
\label{deleq}
\ea
 We utilise the RG-extrapolated values of the running charged fermion  masses 
at the intermediate scale $\mu=M_R \approx v_R \approx 10^{13}$ GeV  as  shown in Table.\ref{tab2} for  $\tan\beta=10, 55$ \cite{prd}. In the present model the definition $\tan\beta = v_u/v_d$ is valid in the presence of MSSM below the intermediate scale.

\begin{table*}
\caption{ Renormalisation Group extrapolated running masses of quarks and charged leptons of three generations
at the intermediate scale $M_{\rm R} = 10^{13}$ GeV as estimated in Ref. ~\cite
{prd}.} 
\begin{ruledtabular}
\begin{tabular}{lcc}\hline
$\tan\beta$ & $10$&$55$\\  
$m_{\rm u}$~(MeV)&$0.888 \pm^{0.169}_{0.179}$&$0.888 \pm^{0.167}_{0.179}$\\    
$m_{\rm c}$~(MeV)&$258.094 \pm^{23.828}_{25.833}$&$258.292\pm^{23.329}_{25.814}
$\\
$m_{\rm t}$~(GeV)&$94.369 \pm^{22.557}_{25.833}$&$104.236 \pm^{32.701}_{18.202}
$\\
$m_{\rm d}$~(MeV)&$1.829 \pm^{0.511}_{0.277}$&$1.821\pm^{0.505}_{0.275}$\\
$m_{\rm s}$~(MeV)&$36.426\pm^{5.158}_{5.480}$&$36.289\pm^{5.077}_{5.434}$\\
$m_{\rm b}$~(GeV)&$1.263 \pm^{0.118}_{0.089}$&$1.576 \pm^{0.264}_{0.168}$\\
$m_{\rm e}$~(MeV)&$0.391 \pm^{0.0002}_{0.0002}$&$0.389 \pm^{0.0005}_{0.0002}$\\
$m_{\mu}$~(MeV)&$82.553 \pm^{0.034}_{0.033}$&$82.206\pm^{0.046}_{0.102}$\\
$m_{\tau}$~(GeV)&$1.408 \pm^{0.0009}_{0.0008}$&$1.657 \pm^{0.018}_{0.014}$\\
\end{tabular}
\end{ruledtabular}
\label{tab2}
\end{table*}

Using the down quark
and charged lepton masses from Table \ref{tab2}  and eqs. (\ref{deleq}) 
we obtain,
\be
\beta_0 ~=~449.773 ~{\rm MeV}, ~\beta_1 ~=~-625.971 ~{\rm MeV}, ~\beta_2 ~=~224.155 ~{\rm MeV},\nonumber\\
\ee
\be
\delta_0 ~=~-15.791~{\rm MeV}, ~\delta_1 ~=~12.334~{\rm MeV}, ~\delta_2 ~=~-8.074~{\rm MeV}.
\label{vbg}
\ee     
Using low-energy values of CKM matrix elements with its phase
$\delta = 60^{\circ}$ and using the renormalization factor $r_N =
 exp[-(y_{top}^2ln(v_R/m_{top})/{16\pi^2}]\simeq 0.86$
leads to the CKM matrix at $\mu=M_R=10^{13}$ GeV,

\par\noindent
\be V_{CKM}=\left[\br{ccc}
0.973852&0.22720&0.00169097-0.00292880i\\
-0.227985-0.000134610i&0.97301-0.000031405i&0.0369880\\ 
0.00675842 -0.002851022i
&-0.0364054-0.000664840i&0.99925\\
\er\right]. \label{ckm}\ee

Defining ${\hat M}_u=diag(m^0_u,m^0_c,m^0_t)$ ,  at first we obtain elements of 
$M_u$ in terms of the running up-quark masses and CKM elements via,

\be
M_u = V^T_{CKM}{\hat M}_u V_{CKM}. \nonumber
\label{vmu}
\ee  

For $\tan \beta = 10$, using eqs.(\ref{mu10}), (\ref{mu126}), (\ref{ckm}), 
and (\ref{vmu})
and Table \ref{tab2},  determines the three parameters $\alpha_i(i=3,4,5)$ while 
three equations are obtained among the other six complex parameters,
~$\alpha_i (i=0,1,2)$ and  $\gamma_i (i=0,1,2)$,

\ba
\alpha_0+\alpha_1+\alpha_2+\gamma_0+\gamma_1+\gamma_2 &=& 369.414{\pm}^{53.550}_{42.847} -i (4.583{\pm}^{1.099}_{0.702}), \nonumber\\
\alpha_0-\alpha_1+\alpha_2+\gamma_0-\gamma_1+\gamma_2 &=& 94204.062{\pm}^
{22517.333}_{14450.556} - i 8.797\times 10^{-6},\nonumber\\
\alpha_0-2\alpha_2+\gamma_0-2\gamma_2  &=& 17.687 {\pm}^{3.085}_{1.832} - i (3.6192
{\pm}^{0.867}_{0.556}), \label{paru1}
\ea

\ba
\alpha_3 &=& -(3423.163{\pm}^{756.854}_{525.630}) - i(62.685{\pm}^{14.983}_{9.616}),\nonumber\\
\alpha_4 &=& 634.282{\pm}^{152.893}_{96.749} - i (268.690{\pm}^{64.225}_{41.215})
,\nonumber\\
 \alpha_5 &=& -(80.018 {\pm}^{11.102}_{9.056}) + i (9.334{\pm}^{2.185}_{1.434}),
\label{paru2}
\ea
\noindent 
where all parameters are in MeV and
the uncertainties in the RHS of these equations reflect the uncertainties  in the low-energy data \cite{prd}. 
It is clear that the set of three eqs.(\ref{paru1}) leaves undetermined three 
complex (six real ) parameters which  provide a very rich structure to the
model. Because of this, the model may be able  to confront the present neutrino data and even the future precision data that may emerge from 
planned and ongoing oscillation experiments. On the other hand, it is also possible 
that the number of parameters may not ensure faithful representation of neutrino data because of highly non-linear nature of the problem emerging from see-saw mechanism.  

In order to examine the efficiency of the model  in representing the neutrino sector, we use the
standard parametrization of the leptonic Pontecorvo-Maki-Nakagawa-Sakata(PMNS)
 mixing matrix,

\begin{eqnarray}
U_{PMNS}=\left[
 \begin{array}{ccc}
 c_{12}c_{13} & s_{12}c_{13} & s_{13}e^{-i\delta}\\
 -c_{23}s_{12}-s_{23}s_{13}c_{12}e^{i\delta} &
 c_{23}c_{12}-s_{23}s_{13}s_{12}e^{i\delta} & s_{23}c_{13}\\
 s_{23}s_{12}-c_{23}s_{13}c_{12}e^{i\delta} &
 -s_{23}c_{12}-c_{23}s_{13}s_{12}e^{i\delta} & c_{23}c_{13}
 \end{array}
 \right]~{\rm diag}(e^{-i\varphi_1/2},e^{-i\varphi_2/2},1),\label{pmns}
\end{eqnarray}
where  $c_{ij}\equiv \cos\theta_{ij}$,
$s_{ij}\equiv\sin\theta_{ij}$, $\delta$ is the Dirac phase and
$\varphi_1,\varphi_2$ are Majorana phases of neutrinos. These phases
have range from $0$ to $2\pi$.

We use experimental data on neutino oscillations within the  $3\sigma$ 
limit \cite{maltoni}:
\begin{eqnarray}
\nonumber &&~~~~~~0.29\leq\tan^2\theta_{12}\leq0.64,\\
\nonumber &&~~~~~~0.49\leq\tan^2\theta_{23}\leq 2.2,\\
\nonumber &&~~~~~~~~~~~\sin^2\theta_{13}\leq0.054,\\
&& 5.2\leq\Delta m^2_{\odot}/{10^{-5} eV^2} \leq 9.8,\\ 
\nonumber &&~~~~~~1.4 \leq \Delta m^2_{atm}/{10^{-3} eV^2}\leq 3.4.
\label{ndata}
\end{eqnarray}

For numerical analysis we exploit the well defined diagonalisation procedure
for complex and symmetric mass matrices,

\ba
U^{\dagger}M_{\nu}U^* &=& {\rm {diag}} ( m_1, m_2, m_3), \nonumber\\
U^{\dagger}M_{\nu}M_{\nu}^{\dagger}U &=& {\rm {diag}} ( m^2_1, m^2_2, m^2_3), 
\label{biu}
\ea
where U is a unitary diagonalising matrix , the light neutrino mass matrix 
$M_{\nu}$ has been defined in  eq.(\ref{massdf}) and $m_i(i=1,2,3)$ are positive mass eigen values.

For the sake of simplicity we reduce the parameters of the model by  treating
 the parameters $\gamma_i(i=0,1,2)$  
as real. Then eq.~(\ref{paru1}) determines six real parameters out of a total 
nine,
including real and imaginary parts of $\alpha_i (i=0,1,2)$ and real 
$\gamma_i (i=0,1,2)$. This choice of parameters implies that the CP-violation has its origin only in the quark sector as reflected in the CKM matrix \cite{ckmcp}.

Thus, in addition to the see-saw scale, we are left with three real parameters 
to fit the neutrino oscillation data on four 
quantities, 
$\Delta m^2_{\odot},\Delta m^2_{atm}, \tan^2\theta_{12}$, and 
$\tan^2\theta_{23}$ and make predictions on $\sin\theta_{13}$,
leptonic Dirac phase ($\delta$) and Majorana phases($\varphi_1, \varphi_2$),
sum of the three light neutrino masses $\Sigma m_i$, the effective matrix element for neutrinoless double beta decay, $<m_{ee}>$, and  the  kinematic
neutrino mass $m_{\beta}$ to be measured in  beta decay where
\ba
<m_{ee}> = |{{\sum }_{i=1}^3}(U_{PMNS}^{ei})^2 m_i|,
 ~~m_{\beta}= { ({\sum}_{i=1}^3}|U_{PMNS}^{ei}|^2 m_i^2)^{1/2}.
\label{beta}
\ea

Equivalently, the three real unknown parameters are defined as,
\ba
   \xi &=& \gamma_0 - 2\gamma_2, \nonumber \\ 
   \eta&=& \gamma_0+\gamma_1+\gamma_2, \nonumber\\
   \zeta&=&\gamma_0-\gamma_1+\gamma_2. \label{nupar}
\ea
Even in the constrained parametrisation of the model , we find that $\xi, ~\eta$ and $\zeta$ are quite efficient in describing
the present neutrino oscillation data.
Some examples of our fit to the data   
and  model predictions are shown in Table \ref{tab3}. 

We find that
the see-saw scale is determined to be $M_N= 3.78\times 10^{13}$ GeV for
hierarchial neutrino masses. The first and the second columns show that for fixed values of  $\eta$ and $\zeta$,
 the parameter $\xi$ is very effective in controlling the value of the solar 
neutrino mixing angle ($\theta_{12}$).
 Within the uncertainties
shown in eq.(\ref{paru1}) and eq.(\ref{paru2}), the predicted reactor mixing angle occurs in the range  $\theta_{13}\simeq 3^{\circ}-  5^{\circ}$ which 
is  within the accessible limit of ongoing and planned experiments
\cite{dctc}. The sum of the three neutrino masses are found to be well within the cosmological bound \cite{hanne}.
The leptonic Dirac phase turns out to be closer to 
$\pi$ with $\delta = 2.9 - 3.1$ radians  and the two Majorana phases are within $5.3-5.7$ radians.
 The predicted values of matrix element for double beta decay and the kinematical mass for beta decay are found to be nearly two orders smaller than  the current experimental bounds \cite{beta,katrin, klap}. Similar conclusion has been also obtained for hierarchial neutrinos 
 with $S_4$ flavor symmetry in the non-SUSY standard model
\cite{hagedorn}.  
        The Jarlskog invariant \cite{jarl} is found to vary between $J_{CP}\simeq 2.95\times 10^{-5}$ 
and $J_{CP}\simeq 10^{-3}$ where the smaller (larger) value depends upon how much
closer (farther) is the Dirac phase ($\delta$) from $\pi$.
We observe that the predictions of this model in the neutrino sector made at the high see-saw scale is to remain stable under radiative corrections when extrapolated to low energies especially since the light neutrino mass eigen values are 
small \cite{mkp6}.     
\begin{table*}
\caption{Fit to the available neutrino oscillation data and predictions 
of reactor mixing angle $\theta_{13}$, leptonic Dirac phase ($\delta$), 
Majorana phases ($\varphi_1, \varphi_2$) and the CP violation parameter $J_{CP}$ in the $SO(10)\times S_4$ model with see-saw scale at $M_R = 3.78\times 10^{13}$ GeV and $\tan\beta=10$}  
\begin{ruledtabular}
\begin{tabular}{lccc}\hline  
$\xi$~(GeV)&$1.025$&$1.100$&$1.235$\\    
$\eta$~(GeV)&$2.137$&$2.137$&$2.400$\\
$\zeta$~(GeV)&$25.529$&$25.529$&$25.700$\\
$m_1$~(eV)&$0.00536$&$0.00596$&$0.00801$\\
$m_2$~(eV)&$0.00920$&$0.00956$&$0.01268$\\
$m_3$~(eV)&$0.05000$&$0.05000$&$0.07860$\\
${\sum}_im_{\rm i}$~(eV)&$0.0645$&$0.0675$&$0.09929$\\
$\Delta m^2_{\odot}$~(eV$^2$)&$6\times 10^{-5}$&$6\times 10^{-5}$&
$9.6\times 10^{-5}$\\
$\Delta m^2_{atm}$~(eV$^2$)&$2.5\times 10^{-3}$&$2.5\times 10^{-3}$&$
3.1\times 10^{-3}$\\
$\sin\theta_{12}$&$0.515$&$0.616$&$0.511$\\
$\sin\theta_{23}$&$0.718$&$0.718$&$0.736$\\
$\sin\theta_{13}$&$0.055$&$0.057$&$0.052$\\
$\delta$(radians)&$3.096$&$3.048$&$3.100$\\
$\phi_1$(radians)&$5.67$&$5.46$&$5.65$\\
$\phi_2$(radians)&$5.59$&$5.39$&$5.65$\\ 
$ J_{CP} $&$2.66\times 10^{-4}$&$6.49\times 10^{-4}$&$2.95\times 10^{-5}$\\       $<m_{ee}>$ (eV)&$0.00646$&$0.00742$&$0.00932$\\
$m_{\beta}$ (eV)&$0.00462$&$0.00516$&$0.00600$\\                             
\end{tabular}
\end{ruledtabular}
\label{tab3}
\end{table*}   
           
\section{VII. Summary and Conclusion }
 
In this work we have addressed the question of possible existence of 
R-parity and Parity conserving left-right gauge theory as an
intermediate symmetry in supersymmetric $SO(10)$ grand 
unified theory with manifest one-loop  unification of the gauge couplings.
We 
found that it is possible to have this intermediate gauge symmetry provided both the left-right gauge theory and SO(10) are
extended to contain $S_4$ flavor symmetry. The particle spectrum needed to 
implement the gauge coupling unification is found to
match into different $G_{2213}\times S_4$ representations leading to exactly vanishing values of two RG coefficients. The Higgs spectrum is also found to
 be consistent with the
mass 
spectra analysis for the $SO(10)\times S_4$ model with fine-tuning of certain 
parameters in the Higgs superpotential. At first ignoring the light 
neutrino mass constraint,
the left-right symmetry breaking  scale is allowed 
to have a wide range of values with $M_R = 5\times 10^9$ GeV to $10^{15}$ GeV
and no tuning of the Majorana coupling is needed in this model  to obtain  desired value of the see-saw scale substantially below the GUT scale.
 
We have carried out analysis of fermion masses and mixings 
using RG extrapolated values of the low-energy data at the intermediate scale
$M_R \simeq 10^{13}$ GeV in SUSY $SO(10)\times S_4$ for the first time.   
Even in the case of  constrained model parametrisation where CP-violation
originates only from the quark sector through CKM matrix,
the model is found to fit all values of quark and lepton 
masses and mixings including very large values of  mixings and very small values of  masses in the neutrino sector. The neutrino oscillation data determines the see-saw scale to be  $M_N \simeq 3.8\times 10^{13}$ GeV for hierarchial neutrinos. 
 Apart from predictions on leptonic CP-violating
parameter, Dirac and Majorana phases, the predicted values of the reactor-neutrino
mixing angle, $\theta_{13}\simeq 3^{\circ} -  5^{\circ}$, are accessible to ongoing and planned long baseline experiments
on neutrino oscillations. 

 It would be interesting to investigate prospects of this model with all complex parameters in type I see-saw and 
the case of experimentally testable quasi-degenerate neutrino spectrum with type II see-saw or a combination of both the type I and type II see-saw models 
 in future works.

\section{ACKNOWLEDGMENT}

\begin{acknowledgments}
 The author thanks  R. N. Mohapatra, K. S. Babu  and A. Ilakovac for  discussion. 
\end{acknowledgments}

\end{document}